# Green ILC Concept: Scenarios toward 2050 Carbon Neutrality in Japan and ILC

Masakazu Yoshioka (Iwate University/Iwate Prefectural University)

**Abstract**

This paper describes Japan's scenario for achieving carbon neutrality by 2050 and the policy that should be adopted by the ILC in Japan in line with that policy. This paper only discusses $CO_2$ emissions during operation, not the lifecycle $CO_2$ emissions of the ILC.

1. **Introduction: $CO_2$ emissions from accelerator facilities**

The earth is currently in a warming trend due to natural cycles (Milankovitch cycle). Although some researchers question the warming caused by anthropogenic factors, I believe that rapid changes are occurring that cannot be explained by natural cycles alone, and I also believe that the world and Japanese government policies that set a goal of achieving carbon neutrality by 2050 are very correct.

Accelerators are powered by electricity, and $CO_2$ is emitted during the production of electricity. In the case of the ILC, peak electricity consumption is about 130 megawatts, resulting in an annual consumption of around 700 million kWh, depending on the operating hours. In Japan, the $CO_2$ emissions per kWh are reported annually by the local electric power company. Electricity is produced from three energy sources: (1) fossil fuels, (2) renewable energy, and (3) nuclear power. For instance, in the region where the ILC candidate site is located (in Japan), the proportionate of renewable energy, including hydroelectric power, is 21%, while nuclear power accounts for zero, and others depend on fossil fuels. As a result, the coefficient per kWh is 480 grams (in FY2021). This value is quite large compared to Europe and the United States. When this value is multiplied by the electricity consumption by the ILC yields an annual $CO_2$ emission of 336 kilotons.

Annually, Japan's Ministry of the Environment conducts a quantitative assessment of each municipality in Japan to ascertain their respective $CO_2$ emissions. According to this assessment, Ichinoseki City, where the ILC candidate site is located, has a value of 871 kilotons in 2018. Based on the $CO_2$ emission factor from 2021, the emissions attributed to the ILC constitute 40% of Ichinoseki's total emissions. However, it's worth noting that the $CO_2$ emissions per kWh of electricity are anticipated to decrease to less than 50% of the current level by the time the ILC operation. The emissions of Ichinoseki City at that time will be considerably smaller, so the discussion here is for reference only.

Figure 1 was given to us by Benno List of DESY. The left side of the figure is the group of so-called developed countries, and it is noticeable that the coefficients for Japan and South Korea are high. On the other hand, Europe has a high ratio of renewable energy sources such as hydro, wind, and solar power. Additionally, France, Switzerland and several other countries rely on nuclear power, resulting in the coefficient for Europe being nearly half that of Japan. The situation in Europe is very different from that in Japan, which is an island nation, because the power grid in Europe is connected across national borders. Japan must create a closed scenario in one country.

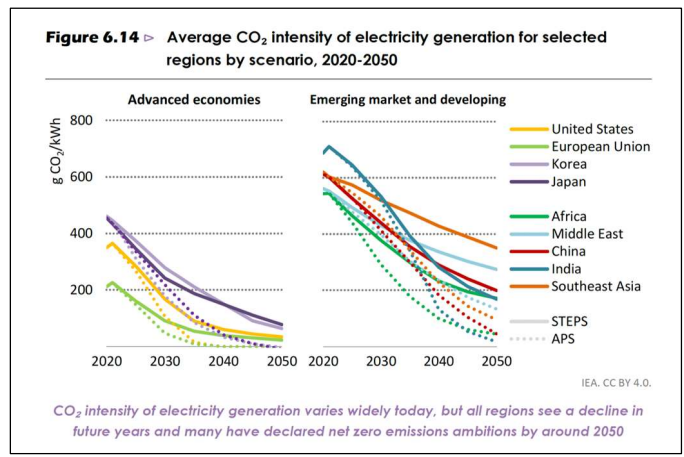

Fig.1 $CO_2$ emission factor/kWh for various countries/regions [1]

What should we do to reduce $CO_2$ emissions? We need to do all three of the following:

1) further develop energy-saving technologies,

2) operate the ILC using as much renewable energy as possible and recover as much of the heat energy emitted from the ILC as possible, and

3) increase $CO_2$ absorption in cooperation with the communities where the ILC will be located. This item 3) will be discussed in detail in the next section.

## 2. Japanese Government's Scenario for Achieving Carbon Neutrality by 2050

According to information provided on the Japanese government's official website, the 2030 target is 37% renewable energy, 21% nuclear, and the rest fossil fuels. The conversion factor is 200 grams, which is about the same as the current European level, according to the Figure 1. But if you look closely, the coefficient for 2050 is not zero, so what happened to the goal of becoming carbon neutral by 2050? According to the government's website, the government will of course make efforts to reduce $CO_2$ emissions to achieve carbon neutrality, but it is not possible to reduce $CO_2$ emissions to zero, and the strategy is to increase $CO_2$ absorption to offset the $CO_2$ emissions.

I find this policy to be realistic. As shown in Figure 2, the Japanese archipelago is located at a moderate latitude, has a mountain spine, the Sea of Japan, and prevailing westerly winds, so rain and snow fall evenly over the country, and the average annual precipitation is 1,700 mm, twice the world average. Agriculture is also carried out using only natural water. Surprisingly, there are many agricultural activities in the world that depends on fossil water. Forests, if properly managed, can be a truly renewable source of energy and absorb and fix $CO_2$. This is called green carbon. In addition, Blue Carbon can be expected to absorb and fix $CO_2$ by cultivating seaweed beds in coastal areas. As an island nation, Japan ranks sixth in the world for the total length of its coastline.

Here we need to know two things. One is "where is carbon stored?" and the other is "the status of biomass that fixes carbon". There was an interesting article in the Asahi Shimbun's GLOBE dated September 21, 2022. The world's greenhouse gas emissions are 52 billion t-$CO_2$ equivalent. On the other hand, carbon is stored in the forests (2 trillion tons) and soil (5.5 to 8.8 trillion tons), of which 3 trillion tons is in the topsoil and 3 trillion tons in the atmosphere. The oceans also store a large amount, but unfortunately the article did not provide that information.

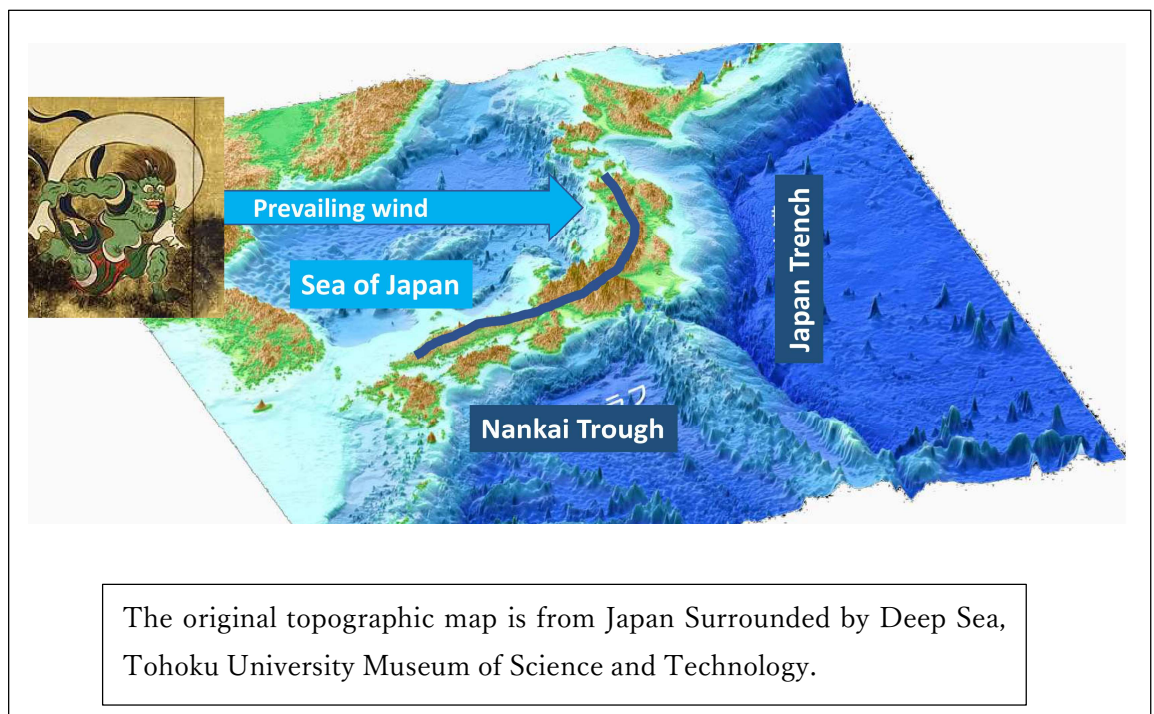

The original topographic map is from Japan Surrounded by Deep Sea, Tohoku University Museum of Science and Technology.

Fig. 2 Why Japan is Blessed with Renewable Biomass

Here is another interesting figure from the paper "The biomass distribution on Earth" by Yinon M. Bar-On, Rob Phillips and Ron Milo [2]. According to this paper, as shown in Figure 3, the total amount of biomass in terms of carbon is 550 gigatons. (The numbers below are in gigatons.) Plants dominate the breakdown at 450 (81%). This is followed by bacteria (70), fungi (12), archaea (7), protists (4), animals (just 2), and viruses (0.2). It is surprising that the biomass of animals is half that of protists. In addition, 50% of the animal biomass is insects, and other interesting information follows, but this is off topic, so I will leave it at that.

Again, Japan is not blessed with fossil fuel resources, but it is tremendously rich in biomass resources due to its mild climate and abundant rainfall.

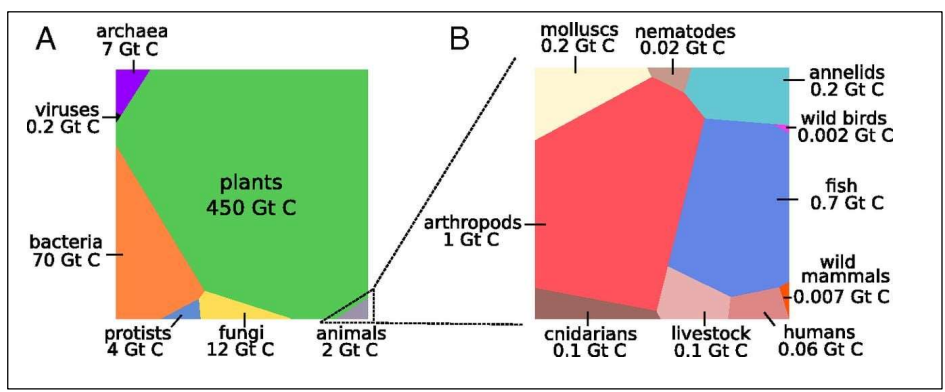

Fig.3 Breakdown of biomass on the earth [2]

Here is a concrete example of a scenario for "achieving carbon neutrality by 2050" in line with the government policy. This is the case of Miyako City, Iwate Prefecture [3]. The $CO_2$ absorption and fixation in the Figure 4 is due to green carbon (forests) and blue carbon (algae).

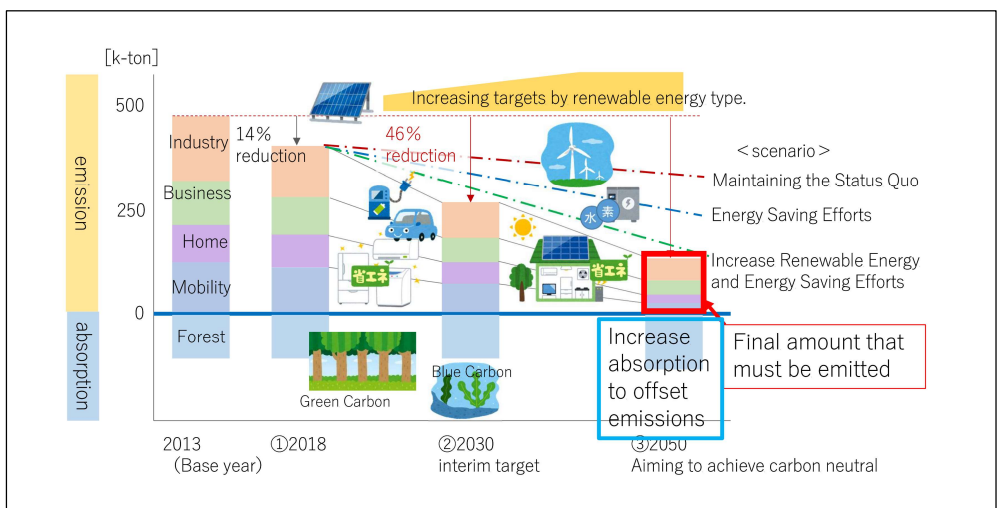

Fig.4 Scenarios for achieving carbon neutrality by 2050 in Miyako City, Iwate Prefecture [3]

### 3. Cooperation with agriculture, forestry, and fisheries is key

What should we take from the above? I read "The role of $CO_2$ absorption should rely on plants, and specifically, "ILC should be linked to the agriculture, forestry, and fisheries industries". Let us look specifically at the case of Ichinoseki City.

We have written before that the ILC will use 700 million kWh of electricity per year. Assuming that the $CO_2$ emission conversion factor is improved to 200 grams/kWh in 2040, the annual $CO_2$ emissions would be 140 kilotons. Of course, I sincerely hope that the ILC will be in full operation by 2040. On the other hand, Ichinoseki City has a forest area of 66,000 hectares, which is quite large for a city of its size. However, natural broadleaf forests cover about half of the area, and properly managed planted forests are not so large. We asked Hiroshi Kikuchi, Regional Forestry Policy Advisor, Ichinoseki City Agriculture and Forestry Department, to provide detailed estimates of the amount of $CO_2$ absorption. The estimation requires a considerable amount of work, since it requires understanding the age, species, and management status of the forest. This is because $CO_2$ absorption increases as trees grow.

The Forestry Agency of Japan estimates that a hectare of a properly managed 40-year-old cedar plantation absorbs 8.8 t of $CO_2$ per year. In reality, this value is not quite reached. According to Kikuchi's estimate, the current $CO_2$ absorption is 300 kilotons per year. Thus, the Ichinoseki City forests absorb an average of 4.5 t $CO_2$ per hectare per year.

This value exceeds the ILC's 2040 $CO_2$ emissions projection, but of course, we must also consider the emissions of our citizens' industries and lifestyles. I would like to emphasize that efforts to visualize numerical targets and link them to concrete action plans are necessary not only for the ILC, but also in other areas.

## 4. Summary

The ILC is expected to be operational in the late 2030s. On the other hand, as the Japanese and global target for achieving carbon neutrality is 2050, it is extremely important that efforts are made to reduce greenhouse gas emissions in the construction and operation of ILC. However, it is almost impossible for ILC facilities to achieve carbon neutrality on its own, and measures to increase and offset $CO_2$ absorption in cooperation with local communities should be considered.

Increasing $CO_2$ absorption can be achieved in cooperation with the agriculture, forestry, and fisheries industries in the region, and it is a future challenge to visualize the figures for both $CO_2$ emissions and absorption, and to formulate concrete action plans.

### Acknowledgment

This paper is based on joint research with many people. In particular, we worked closely with many academies, companies, and basic municipalities in the Tohoku region that are aiming to achieve carbon neutrality by 2050. I would like to express my gratitude to all my collaborators.

## References

[1] IEA (2022), World Energy Outlook 2022, IEA, Paris, https://www.iea.org/reports/world-energy-outlook-2022, ,
ssLicense: CC BY 4.0 (report); CC BY NC SA 4.0 (Annex A)

[2] "The biomass distribution on Earth" by Yinon M. Bar-On, Rob Phillips and Ron Milo, https://www.google.com/search?client=firefox-b-d&q=The+biomas+distribution+on+earth

[3] From the website of Miyako City, Iwate Prefecture
https://www.city.miyako.iwate.jp/energy/r4miyakoshi_renewableenergy_promotionplan.html (Japanese only)